\documentclass[12pt,a4paper]{article}
\usepackage{amsmath}
\usepackage{graphicx}
\usepackage{verbatim} 
\usepackage{setspace}
\newcommand{\ddt}{\frac{\partial}{\partial t}}
\newcommand{\ddr}{\frac{\partial}{\partial r}}

\begin{document}
\title{Discrete Self-Similarity in Ultra-Relativistic Type-II Strong Explosions}

\author{Yonatan Oren \\ \small{Racah Institute of Physics, the Hebrew University, 91904 Jerusalem, Israel} \\
       Re'em Sari \\ \small{ Racah Institute of Physics, the Hebrew University, 91904 Jerusalem, Israel} 
                  \\ \small{and California Institute of Technology, MC 350-17, Pasadena, CA 91125 }}
\date{}   

\maketitle

\begin{abstract}
A solution to the ultra-relativistic strong explosion problem with a non-power law density gradient is delineated. We consider a blast wave expanding into a density profile falling off as a steep radial power-law with small, spherically symmetric, and log-periodic density perturbations. We find discretely self-similar solutions to the perturbation equations and compare them to numerical simulations. These results are then generalized to encompass small spherically symmetric perturbations with arbitrary profiles.
\end{abstract}

\section{Introduction}
The profusion of explosions occurring in the observable universe has led to a pointed interest in the dynamics of blast waves. The quantitative treatment of strong explosions began with the self-similar solutions found by Sedov, Von-Neumann and Taylor \cite{Sedov} \cite{VonNeumann} \cite{Taylor} for the flow behind spherical Newtonian shocks propagating into a cold gas with a power-law density profile, $\rho \propto r^{-k}$. This solution is valid for moderately steep decay exponents, $k<3$. Subsequently, corresponding solutions were found in the ultra-relativistic regime by Blandford \& McKee \cite{BM} for $k<4$. In these solutions the Lorentz factor of the shock $\Gamma$ scales as $\Gamma^2 \propto t^{-m}$, and $m$ is fixed by energy conservation arguments. For $k>4$ this procedure fails due to the energy in the Blandford-McKee solution diverging, and a different argument must be used if a self-similar solution is to be found. It turns out that there exists a $k_g>4$ such that for $k>k_g$ just such an argument exists \cite{Best}. The solutions in this regime are called type-II solutions \cite{Zeldovich}, and what sets them apart from the solutions with $k<4$, known as type-I solutions, is the rapid acceleration of the shock and the fluid behind it that causes the formation of a sonic point between the shock and the center of the explosion. This point (actually a spherical surface) marks the boundary of an inner region that becomes causally disconnected from the shock. This allows a self-similar flow behind the shock to coexist with a non self-similar flow further away from it, thus maintaining a finite amount of total energy. For the ultra-relativistic case, $k_g=5-\sqrt{3/4} \approx 4.13$. The sonic point appears as a singularity in the hydrodynamic equations, and the requirement of regularity in traversing this singularity supplies the necessary condition to fix the  value of $m$.

In this paper we focus on Type-II ultra-relativistic solutions with $k>k_g$ \cite{Best}, and use these as the basis for a perturbative analysis where we disturb the external density profile. The basic method for doing so was developed in a previous paper \cite{Oren} for the Newtonian case, and we adapt it here for the ultra-relativistic case. We introduce a special family of perturbations to the ambient medium surrounding the explosion, so that we are able to reduce the hydrodynamic equations to a set of ordinary differential equations, and then find the flow behind the shock in the presence of perturbations. In section \ref{sec:unpert} we briefly describe the unperturbed solutions, while in section \ref{sec:RI} we shed light on some general properties of these solutions. In section \ref{sec:pert} we write down the equations for the perturbations and solve them, and in section \ref{sec:fourier} we generalize the results of the previous section by using a spectral decomposition of an arbitrary perturbation profile. Finally in section \ref{sec:discuss} we summarize and discuss our results.

\section{The Unperturbed Solution}\label{sec:unpert}
We begin with a quick review of the self-similar solutions that will later serve as the basis for perturbation. These describe the flow after the discharge of a large amount of energy in a small volume surrounded by a spherically symmetric distribution of stationary cold gas. The density of this medium follows a radial power law, 
\begin{equation} \label{EQ:rho}
\rho(r) = K r^{-k}.
\end{equation} 
A more detailed treatment is given in Best \& Sari \cite{Best}. We use units such that the speed of light is unity throughout the paper.

\subsection{The hydrodynamic equations}
The hydrodynamic equations for a relativistic ideal gas are derived from the local conservation of energy, momentum and number of particles. In spherical symmetry they take the form
\begin{align} \label{EQ:hydro}
&\ddt \left[ \frac{e+\beta^2 p}{1-\beta^2} \right]+\frac{1}{r^2} \ddr \left[ r^2\frac{(e+p)\beta}{1-\beta^2} \right] = 0 \nonumber \\
&\ddt \left[ \frac{(e+p)\beta}{1-\beta^2} \right]+\frac{1}{r^2} \ddr \left[ r^2\frac{(e+p)\beta^2}{1-\beta^2} \right] 
      +\frac{\partial p}{\partial r} = 0 \nonumber \\
&\frac{\partial n'}{\partial t}+\frac{1}{r^2} \ddr \left[ r^2 n' \beta \right] =0   
\end{align}
where the velocity $\beta$ and the particle number density $n'$ are measured in the frame of the unshocked gas, and the pressure $p$ and internal energy density $e$ are measured in the fluid rest frame. The particle density in the fluid rest frame is related by $n'=\gamma n$ to that in the explosion frame, with $\gamma=(1-\beta^2)^{-1/2}$. We will treat these equations in the ultra-relativistic limit where $\Gamma>>1$, and use the relativistic equation of state: 
\begin{equation} \label{EQ:EOS}
p=\frac{1}{3} e.
\end{equation}

We take the Lorentz factor of the shock to scale as 
\begin{equation} \label{EQ:G(t)}
\Gamma^2 \propto t^{-m}
\end{equation}
so the shock radius is, to first order in $1/\Gamma^2$, 
\begin{equation}
R(t)=t \left[ 1-\frac{1}{2(m+1)\Gamma^2} \right],
\end{equation}
where $m$ is as yet undetermined. We now make a change of variables to a similarity coordinate that follows the scale height in the problem, $R/\Gamma^2$. The coordinate $\chi$ is defined by
\begin{equation} \label{EQ:chi}
\chi=[1+2(m+1)\Gamma^2](1-r/t).
\end{equation}
We define the self-similar functions $f$, $g$ and $h$ as
\begin{align} \label{EQ:SSfunctions}
p        &=\frac{2}{3} w_1 \Gamma^2 f(\chi) \nonumber \\
\gamma^2 &= \frac{1}{2} \Gamma^2 g(\chi) \nonumber \\
n'       &=2 n_1 \Gamma^2 h(\chi)
\end{align}
where we can take $w_1$, the enthalpy before the shock, equal to $\rho_1$ because the unshocked gas is assumed to be cold.

Substituting equations \eqref{EQ:chi} and \eqref{EQ:SSfunctions} into equation \eqref{EQ:hydro} we get the hydrodynamic equations in their self-similar form:
\begin{align} \label{EQ:unpert}
f'(\chi) &= gf  \frac{4[2(m-1)+k]-(m+k-4)g\chi}{(m+1)(4-8g\chi+g^2 \chi^2)} \nonumber \\
g'(\chi) &= g^2 \frac{(7m+3k-4)-(m+2)g\chi}{(m+1)(4-8g\chi+g^2 \chi^2)} \nonumber \\
h'(\chi) &= gh  \frac{2(9m+5k-8)-2(5m+4k-6)g\chi+(m+k-2)g^2 \chi^2}{(m+1)(2-g \chi)(4-8g\chi+g^2 \chi^2)}.
\end{align}
The boundary conditions for these equations are derived from the Rankine-Hugoniot conditions at the shock which  can be written as
\begin{align} \label{EQ:BC}
p_2 &=\frac{2}{3} \Gamma^2 w_1 \nonumber \\
n_2'&=2 \Gamma^2 n_1 \nonumber \\
\gamma_2^2&=\frac{1}{2} \Gamma^2,
\end{align}
where subscript $1$'s denote quantities just ahead the shock and $2$'s quantities immediately behind the shock.  
In terms of the self-similar quantities, these boundary conditions become
\begin{equation} \label{EQ:BCunpert}
f(\chi=1)=g(\chi=1)=h(\chi=1)=1.
\end{equation}
The self-similar form \eqref{EQ:unpert} of the hydrodynamic equations exposes the singularity at the sonic point, $g(\chi_s) \chi_s=2(2-\sqrt{3})$, where the denominators vanish. A valid solution should be regular everywhere, including the sonic point, and so we can require the numerators as well as the denominators to vanish there. This condition yields the proper value of $m$,
\begin{equation}\label{EQ:m}
m=k(3-2\sqrt{3})-4(5-3\sqrt{3})
\end{equation}
The equations \eqref{EQ:unpert} may be solved implicitly in terms of the variable 
\begin{equation}\label{EQ:x}
x=\frac{g\chi +10\sqrt{3}-2\sqrt{3}k-4}{10\sqrt{3}-2\sqrt{3}k-3}
\end{equation}
to give
\begin{align} \label{EQ:unpert_sol}
log(g)&=(k-6)(3-2\sqrt{3})log(x) \nonumber \\
log(f)&=(k-6)(4-2\sqrt{3})log(x) \nonumber \\
log(h)&=\frac{(k-6)[33-20\sqrt{3}+(4\sqrt{3}-6)k]log(x)+(k+2\sqrt{3}-4)log(2-g\chi)}{1-5\sqrt{3}+\sqrt{3}k}.
\end{align}
These solutions now form the basis of the perturbative analysis to be discussed in section \ref{sec:pert}.

\section{Type-II solutions as simple waves} \label{sec:RI}
The hallmark of Type-II flows is the existence of a sonic point. We now digress to explore a special property that arises from the combination of a sonic point with ultra relativistic self similar flows. We begin by considering the Riemann invariants (RI) of the flow \cite{Landau}. Simply put, the RI are quantities conserved along characteristics of the flow. Considering only the $C_\pm$ characteristics here, this condition may be written as an advection equation for the respective invariants $J_\pm$:
\begin{equation} \label{EQ:RIadvect}
\left[ \frac{\partial}{\partial t} + \frac{u \pm c}{1 \pm u c}  \frac{\partial}{\partial r} \right] J_\pm \equiv D_\pm J_\pm = 0.
\end{equation} 
where $c$ is the local speed of sound and $D$ is defined as the advective derivative. For the case  of plane waves, the RI are given by \cite{Johnson}
\begin{equation} \label{EQ:RIplanar}
J^{planar}_\pm = \int \frac{dp}{c(e+p)} \pm \frac{1}{2} ln \left( \frac{1+\beta}{1-\beta} \right),
\end{equation}
Which evaluates under the relativistic equation of state \eqref{EQ:EOS} to
\begin{equation}
J^{planar}_\pm = \frac{\sqrt{3}}{4} ln(p) \pm \frac{1}{2} ln (4\gamma^2).
\end{equation}

However, it may be seen by substitution in equation \eqref{EQ:hydro} that for spherical geometry these quantities satisfy 
\begin{equation} \label{EQ:RIspherical}
D_\pm J^{planar}_\pm = - \frac{2}{r} \frac{uc}{1+uc}.
\end{equation} 

The right hand side of \eqref{EQ:RIspherical} must be incorporated into the advective term on the left for us to find the spherical $J_\pm$ that satisfy \eqref{EQ:RIadvect}. This is in general fruitless, but we may here take into account the nature of relativistic self-similar flows, and find a solution valid under the ultra-relativistic approximation prevalent throughout this work. As we saw earlier, the solutions at hand are concentrated on a thin shell with thickness of order $R/\Gamma^2$. That means that in the limit of high Lorentz factors we may safely take $r \approx R \approx t$, and substitute that into \eqref{EQ:RIspherical}. Further taking $\beta \approx 1$ and $c=1/\sqrt{3}$, we arrive at
\begin{equation}
D_\pm \left( J^{planar}_\pm + \frac{2}{\sqrt{3}+1} ln(t) \right) = 0.
\end{equation}

so we may write
\begin{equation} \label{EQ:RI}
J_\pm = \gamma^2 p^{\pm \sqrt{3}/2} t^{\pm 4/(\sqrt{3}+1)}.
\end{equation}

\begin{figure} 
\includegraphics[scale=0.3]{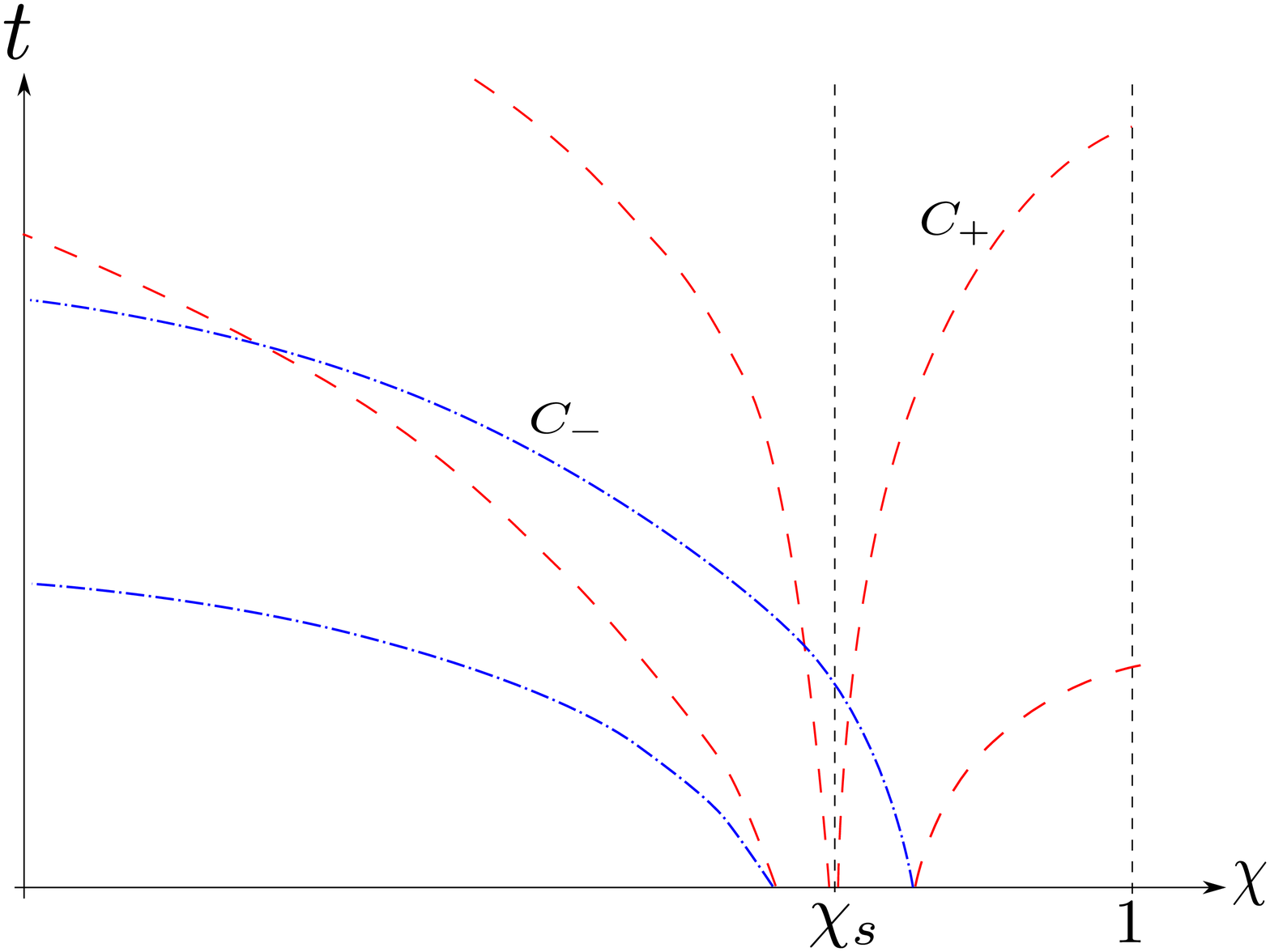}
\caption{The causal structure of Type-II solutions. At late times all of space is covered by $C_+$ characteristics that originate at the sonic point, making $J_+$ constant. }
\label{fig:causal}
\end{figure}

Plugging in the solutions \eqref{EQ:unpert_sol} into \eqref{EQ:RI} reveals that $J_+$ is a constant (both along $C_+$ characteristics and across them). This is somewhat surprising because there is no reason for either of the RI to be constant in a general flow. We can understand how it comes about by considering the special causal structure created by the presence of a sonic point. The sonic point acts as a stationary point for outgoing $C_+$ characteristics, not unlike the event horizon for outgoing light rays in a gravitational black hole. Thus we can see, as shown in figure \ref{fig:causal}, that all $C_+$ characteristics asymptotically approach the sonic point when traced back to early times. This means that the $C_+$ characteristics that eventually fan out to cover all the space below the shock  carry with them a single value of $J_+$, the one at the sonic point.

\begin{figure} 
\includegraphics[scale=0.5]{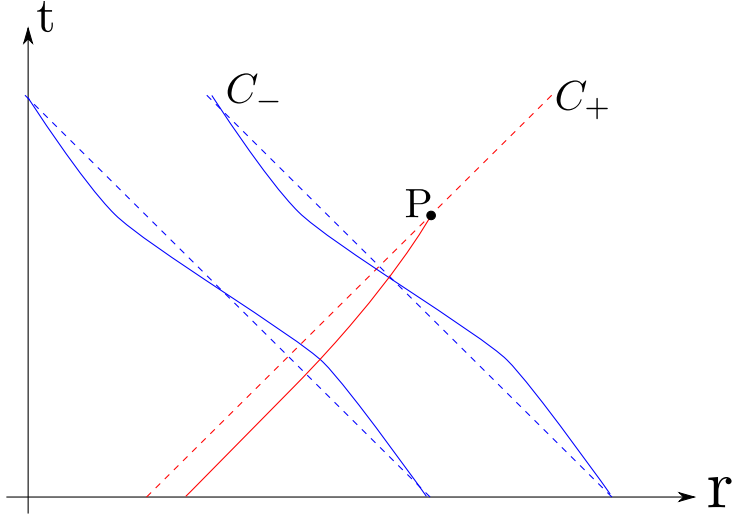} 
\caption{The two sets of characteristics and their interaction. Solid and dashed lines represent perutbed and unperturbed characteristic curves, respectively. In this example perturbations to $J_-$ create a deflection of the $C_+$ characteristics, thus causing a different characteristic to reach the point $P$. This however will not make a difference at $P$ if $J_+$ is constant.}
\label{fig:characteristics}
\end{figure}

Flows with a constant value of one RI are called simple waves. The crucial consequence of this property is that sound waves traveling along the direction opposite to that associated with the constant RI (ingoing waves in our case) will do so without being scattered, even though they are traversing a non uniform and non steady background. This can be understood by considering the role of RI in wave scattering. If we consider (without loss of generality) left going sound waves in a general flow, these waves will create perturbations to the fluid velocity and to the speed of sound. These in turn will cause the right going characteristics to be slightly deflected, and thus the value carried by $J_+$ to some point $P$, as depicted in figure \ref{fig:characteristics}, will deviate from the same value without the left going waves, giving rise to scattered right going waves. However, if $J_+$ is constant, and the hydrodynamic variables are definite functions of the RI, this has no effect. The same value of $J_+$ will arrive at $P$, and no change in the hydrodynamics will be felt there.

The latter condition is satisfied in our case by virtue of the relativistic equation of state. If a general relation exists between $p$, $e$ and $\rho$ the integral in \eqref{EQ:RIplanar} will be path dependent and thus hydrodynamic variables at $P$ will depend both on the value of $J_+$ and on the history of the $C_+$ characteristic that arrived there, nullifying our previous result. We now come to the conclusion of this section, where we see that several effects conspired, in the case discussed here, to make our flow a simple wave and thus greatly simplify the perturbation analysis.

\section{Perturbations}\label{sec:pert}
Applying an arbitrary perturbation to the ambient density profile \eqref{EQ:rho} will in general lead to a non self-similar flow that will be difficult to solve for without a full numerical simulation. We can however carefully choose a special perturbation that results in tractable equations for the perturbed flow. We consider a log-periodic perturbation,
\begin{equation} \label{eq:rhopert}
\rho+\delta \rho=K r^{-k} \left[ 1+\sigma \left( \frac{r}{r_0} \right) ^{i\beta} \right].
\end{equation}
The merit of this choice is that the perturbation wavelength scales like the radius, introducing no new scale into the problem. The parameter $\sigma$ is a small dimensionless amplitude, and $\beta$ is the logarithmic frequency. $r_0$ has dimensions of length and only serves to determine the phase of the perturbation. While the physical perturbation is only the real part of  $\delta \rho$ in \eqref{eq:rhopert}, writing it as a complex power law facilitates finding a self-similar solution. We rewrite the perturbed hydrodynamic functions and shock radius as 
\begin{align} \label{EQ:def_pert}
p        &=\frac{2}{3} w_1 \Gamma^2 [ f(\chi) +b(t) \delta f(\chi) ]\nonumber \\
\gamma^2 &= \frac{1}{2} \Gamma^2 [ g(\chi) +b(t) \delta g(\chi) ] \nonumber \\
n'       &=2 n_1 \Gamma^2 [ h(\chi) +b(t) \delta h(\chi) ]
\end{align}
and 
\begin{equation}
R+\delta R = t \left[ 1-\frac{1-b(t)}{2(m+1)\Gamma^2} \right]
\end{equation} 
where we take $b(t)$ to be
\begin{equation} \label{EQ:b}
b(t)=\frac{\sigma}{d}R^{q}.
\end{equation}
We now have two unknown parameters: $q$, that describes the frequency of the perturbations behind the shock, and $d$ which describes their complex amplitude. Of these, $q$ may be readily seen from the boundary conditions to satisfy $q=i\beta$. As can be expected in a linear perturbation analysis this means that the perturbations behind the shock oscillate at the same frequency as the disturbance before the shock. The value of $d$, which connects the strength of the perturbations ahead of and behind the shock, is less obvious and must be solved for along with the perturbations. We can now substitute \eqref{EQ:def_pert} into \eqref{EQ:hydro} and linearize with respect to $b(t)$ to obtain equations for $\delta f$, $\delta g$ and $\delta h$. These equations may be written as 
\begin{equation} \label{EQ:SSpert}
M Y' + L Y = 0, 
\end{equation} 
where $Y(\chi)=(\delta f,\delta g,\delta h)^T$, 
\begin{align}\label{EQ:M}
M=\frac{1}{m+1}\begin{bmatrix} 
  	g(g\chi-4)	& 4f 		 		& 0					\\
  	g\chi-1		& f\chi    			& 0					\\
    0			& \frac{2h}{g^2}	& \chi-\frac{2}{g} 
  \end{bmatrix}, \nonumber \\
\end{align}
and
\begin{align}  \label{EQ:L}
  L=\begin{bmatrix} 
  	(k+m-q-4)g^2	& 2(k+m-4)fg 			& 0				\\ 
  	(k+2m-q-2)g 	& (k+2m-q-2)f   		& 0				\\ 
    0				& 2(k+m-2)\frac{h}{g}	& k+m-q-2.
  \end{bmatrix} \nonumber \\
  +(m+1)
  \begin{bmatrix} 
  	4g'	 	& 2(g\chi-2)f'				& 0				\\ 
  	g' \chi & f' \chi   				& 0				\\ 
    0		& 2(g\chi -1)\frac{h'}{g^2}	& 2\frac{g'}{g^2}.
  \end{bmatrix}
\end{align}

It is notable that the equations for the pressure and velocity are decoupled from the density, a consequence of the relativistic limit where the rest mass is a negligible part of the energy. These equations must be solved in conjunction with the proper boundary conditions. The conditions at the shock are derived from equation \eqref{EQ:BC} by comparing the perturbed functions at the location of the unperturbed shock with the unperturbed functions at the same point, to first order in $b$. This yields
\begin{align} \label{EQ:BCshock}
\delta f(\chi=1) &= \frac{q+m+1}{m+1}+d+f'(1) \nonumber \\
\delta g(\chi=1) &= \frac{q+m+1}{m+1}+g'(1) \nonumber \\
\delta h(\chi=1) &= \frac{q+m+1}{m+1}+d+h'(1).
\end{align}
One additional condition is needed to close this system and determine the value of $d$, and that is the boundary condition at the sonic point. The condition of regularity at the sonic point is again satisfied if the numerators in equation \eqref{EQ:SSpert} vanish at the sonic point. This yields the condition
\begin{equation} \label{EQ:BCsonic}
\frac{\delta f(\chi_s)}{f(\chi_s)} = -\frac{2}{\sqrt{3}} \frac{\delta g(\chi_s)}{g(\chi_s)},
\end{equation}
which together with equations \eqref{EQ:SSpert} and \eqref{EQ:BCshock} enables us to find both $d$ and the functions $\delta f$, $\delta g$ and $\delta h$ for any $k$ and $\beta$. The straightforward solution would be to start with \eqref{EQ:BCshock} at the shock and use a shooting method to find the value of $d$ that satisfies equation  \eqref{EQ:BCsonic} at the sonic point. 

We can, though, use the insight provided by the discussion in section \ref{sec:RI} to simplify the problem. The shock acts as a source of ingoing sound waves that propagate as the perturbations to $u$ and $p$ that we are seeking. As we have shown in section \ref{sec:RI} the unperturbed solutions is a simple wave, and so these ingoing waves travel without scattering towards decreasing radii. Beyond the sonic point all characteristics point at decreasing values of $\chi$, and thus no reflections may travel back out. This is true wherever the self-similar solution is valid, which is our scope of interest. This region covers a fraction of the volume inside the shock that asymptotically approaches unity. We thus see that to solve the perturbations equations we need only take into account ingoing waves. The form of $J_+$ may be differentiated to reveal the connection between pressure and velocity perturbations:
\begin{equation} \label{EQ:leftwave}
\frac{\delta p}{p}=-\frac{2}{\sqrt{3}} \frac{\delta \gamma^2}{\gamma^2}.
\end{equation}

\begin{figure} 
\includegraphics[scale=0.4]{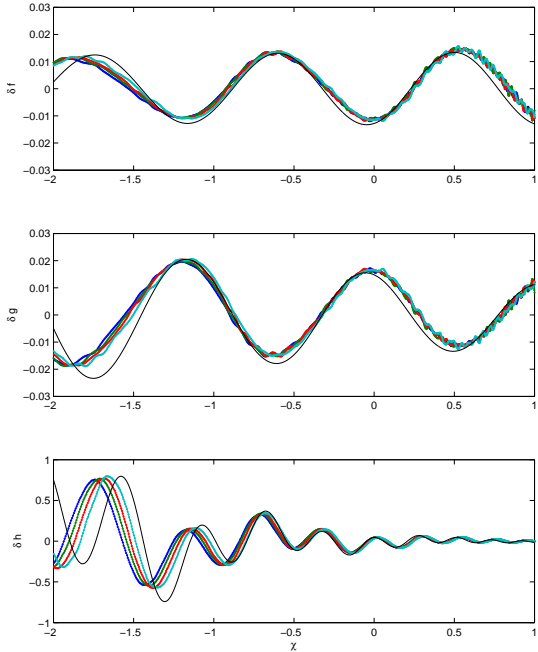}
\caption{The solution to the perturbation equations against the result of a full non-linear numerical simulation. The solid black line is the theoretical value calculated from equation \eqref{EQ:SOL}, and the dotted lines in color represent the numerical solution, converted to self similar form using equation \eqref{EQ:def_pert}. The four lines represent four consecutive periods of the external perturbation, demonstrating the dicretely self-similar nature of the solutions.}
\label{fig:dgfh}
\end{figure}

Equation \eqref{EQ:BCsonic} therefore holds everywhere and not only at the sonic point. This relation enables us to find $\delta g$, $\delta f$ and $d$ analytically:
\begin{align} \label{EQ:SOL}
\delta g &= (2\sqrt{3}+3)\frac{2-2m-k+i\beta}{m+1}
       \frac{\sqrt{1+\frac{12}{\sqrt{3}g\chi-4\sqrt{3}-6}}}{\sqrt{4-8g\chi+g^2\chi^2}} 
       x^{3+m-i\beta} \nonumber \\
\delta f &= -\frac{2}{\sqrt{3}} f \frac{\delta g}{g} \nonumber \\
d        &= -\left(1+\frac{2}{\sqrt{3}}\right) \left(1+\frac{i\beta}{m+1}\right).
\end{align}
We are left only with the equation for $\delta h$ that may be solved using numerical methods. The value of $d$ reveals that at large frequencies the amplitude of perturbations behind the shock is inversely proportional to the frequency of the density perturbations $\beta$, and the phase is a quarter wave behind them. Representative solutions are shown in figure \ref{fig:dgfh} along with corresponding results of a full numerical simulation performed with a second order Godunov type scheme.  
 
It bears mention that while $f$, $g$ and $h$ are functions of $\chi$ only, the solution we find is not strictly self-similar. It is rather composed of one truly self-similar part, the unperturbed solution, and another part which is discretely self-similar, the perturbation. While the perturbation appears to be in itself continuously self-similar, this is only because we used complex notation to write it. Once we take the real part of the complex exponential in \eqref{EQ:b} the oscillatory nature of the perturbations is revealed and we get only discrete self-similarity, with a period of 
\begin{equation} \label{EQ:period}
\frac{\delta R}{R} = e^{\frac{2\pi}{\beta}}-1,
\end{equation}
very much like the Newtonian case treated in Oren \& Sari \cite{Oren}.

\section{Arbitrary Perturbations} \label{sec:fourier}
\begin{figure} 
\includegraphics[scale=0.4]{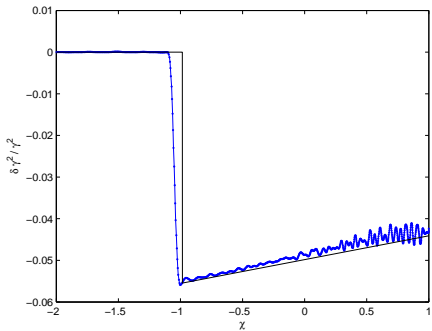}
\caption{The relative Lorentz factor perturbation following a sudden increase in the ambient density, with $k=5$, $s=0.05$, $r_j=10$ and $R/r_j=1.75$. The numerical solution is smoothed to reduce numerical noise.}
\label{fig:step}
\end{figure}

The method outlined so far is limited by the special choice of density perturbations. However, since these special  perturbations form a complete set we can use them to decompose any given perturbation. The linearity of our scheme ensures that summing the solutions with the appropriate weights will give us a solution for the compound perturbation, in a similar fashion to the well known Fourier method for solving differential equations. The resulting solution will no longer be self-similar, due to the different time dependence of each component. We demonstrate the validity of this procedure by considering a scenario commonly encountered in astrophysical context, an abrupt density jump:

\begin{equation}
\rho(r)=K r^{-k} [1+s \theta(r-r_j)]
\end{equation}
where $\theta(x)$ is the Heaviside step function, $r_j$ is the location of the discontinuity and $s$ is a small amplitude. In figure \ref{fig:step} we compare the solution obtained by a fourier decomposition of the step function to a full numerical simulation with the corresponding initial conditions, at a specific point in time. The discontinuity depicted there is a reverse shock propagating back into the shocked material and slowing it down.  
  
\section{Discussion} \label{sec:discuss}
\begin{figure} 
\includegraphics[scale=0.5]{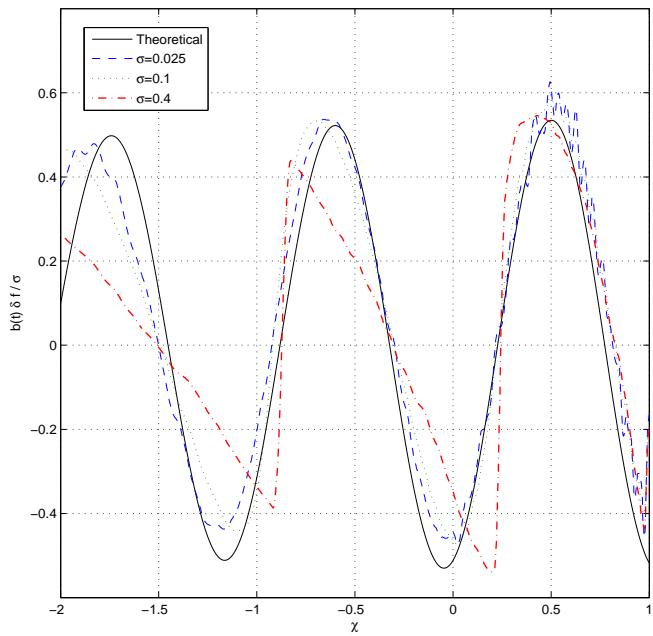}
\caption{The pressure perturbation for $k=5$, $\beta=20$ and different values of $\sigma$, normalized by $\sigma$ to facilitate comparison. The solid black line represents the analytic solution. }
\label{fig:sigmas}
\end{figure}

We have laid out a method of solving the hydrodynamic equations for a strong explosion in the presence of spherically symmetric perturbations to the ambient density. At first we study a special group of perturbations with log-sinusoidal radial dependence, and discover an analytic solution. The requirement of spherical symmetry is not easily relaxed, because relativistic effects make it difficult to find self-similar solutions with a non trivial angular dependence. Another limitation is the linearity of the perturbation analysis that  limits the validity of the solutions to small amplitudes. The smallness required may be seen in figure \ref{fig:sigmas} where numerical solutions with different amplitudes are superimposed against the analytical solution. It can be seen that our theory gives reasonably accurate results for amplitudes up to about $0.1$. The nonlinearity of waves with higher amplitudes is expressed through shock formation before their crests, as can be seen in the red line in figure \ref{fig:sigmas}.

On the other hand we take advantage of linearity to generalize our results using a Fourier-like method, decomposing an arbitrary perturbation to simple modes for which we can solve the equations analytically. In this way we can treat interesting scenarios like a sudden rise or drop in density, which might e.g. be encountered in a stellar wind due to interaction with the interstellar medium. Another possible application is the emergence of a shock from the edge of a star \cite{Pan}, where the drop in density accelerates shocks to relativistic velocities. The effect of perturbations in the star's envelope can be treated with the method presented here, requiring only an adaptation of the unperturbed solution.

Acknowledgments: The authors wish to thank Prof. Tsvi Piran for fruitful discussions. This research was partially 
supported by a NASA grant, IRG grant and a Packard Fellowship.

\pagebreak

\end{document}